\documentclass[
 aip,
 jcp,%
 amsmath,amssymb,
preprint,%
]{revtex4-1}

\usepackage{graphicx}
\usepackage{dcolumn}
\usepackage{bm}
\usepackage[T1]{fontenc}
\usepackage[utf8]{inputenc}
\usepackage{lmodern}
\usepackage{graphicx,xcolor}
\usepackage{hyperref}
\usepackage{amsfonts}
\usepackage{amsmath}
\usepackage{amssymb}
\usepackage{amsthm}
\usepackage{amssymb}
\usepackage{mathrsfs}
\usepackage{epstopdf}
\usepackage{braket}
\usepackage{tikz}
\usepackage{xifthen}                                               
\usetikzlibrary{positioning} 
\usepackage{chemfig}
\usepackage{graphicx,xcolor}
\usepackage{setspace}
\usepackage{soul}
\begin{document}

\title{Sign learning kink-based (SiLK) quantum Monte Carlo for molecular systems}

\author{\surname{Xiaoyao} Ma}
\affiliation{Department of Physics and Astronomy, Louisiana State University, Baton Rouge, LA 70803, USA}
\author{\surname{Randall W} Hall}
\affiliation{Department of Natural Sciences and Mathematics, Dominican University of California, San Rafael, CA 94901, USA}
\affiliation{Department of Chemistry, Louisiana State University, Baton Rouge, LA 70803, USA}
\author{\surname{Frank} Löffler}
\affiliation{Center for Computation \& Technology, Louisiana State University, Baton Rouge, LA 70803, USA}
\author{\surname{Karol} Kowalski}
\affiliation{William R. Wiley Environmental Molecular Sciences Laboratory, Battelle, Pacific Northwest National Laboratory, Richland, Washington 99352, USA}
\author{\surname{Kiran} Bhaskaran-Nair}
\affiliation{Department of Physics and Astronomy, Louisiana State University, Baton Rouge, LA 70803, USA}
\affiliation{Center for Computation \& Technology, Louisiana State University, Baton Rouge, LA 70803, USA}
\author{\surname{Mark} Jarrell}
\affiliation{Department of Physics and Astronomy, Louisiana State University, Baton Rouge, LA 70803, USA}
\affiliation{Center for Computation \& Technology, Louisiana State University, Baton Rouge, LA 70803, USA}
\author{\surname{Juana} Moreno}
\affiliation{Department of Physics and Astronomy, Louisiana State University, Baton Rouge, LA 70803, USA}
\affiliation{Center for Computation \& Technology, Louisiana State University, Baton Rouge, LA 70803, USA}

\date{\today}

\begin{abstract}
The Sign Learning Kink (SiLK) based Quantum Monte Carlo (QMC) method is
used to calculate the \textit{ab initio} ground state energies for multiple geometries of the H$_{2}$O, N$_2$,
and F$_2$ molecules. The method is based on Feynman's path integral formulation of quantum mechanics and has two
stages. The first stage is called the learning stage and reduces the well-known QMC minus sign problem by
optimizing the linear combinations of Slater determinants which are used in the second stage,  a conventional QMC simulation. The 
method is tested using different vector spaces and compared to the results of
other quantum chemical methods and to exact diagonalization.  Our findings demonstrate
that the SiLK method is accurate and reduces or eliminates the minus sign problem.
\end{abstract}

\pacs{05,31,82}
\maketitle

\section{Introduction}
The development of accurate and computationally tractable \textit{ab initio} methods for studying correlated 
electronic systems ranging from single molecules to bulk materials~\cite{dreuw2005single} is an area of wide 
interest.  Feynman's path integral formulation of quantum mechanics~\cite{feynman2005quantum} has long attracted 
attention due to its ability to include exact correlation and  finite temperature effects, as well as providing 
a method that can simultaneously treat electronic and geometric degrees of freedom.  The path integral formulation 
is one of a number of methods commonly referred to as Quantum Monte Carlo (QMC)-based 
algorithms~\cite{foulkes2001quantum}.

In general, the use of QMC-based algorithms  are hindered by  the so-called minus sign problem in which the 
fluctuating sign of the fermionic density matrix leads to statistical errors that scale exponentially with 
inverse temperature and system size.  The minus sign problem~\cite{troyer2005computational,loh1990sign} 
remains a great challenge in condensed matter physics and quantum chemistry.  

In quantum chemistry, there are a number of methods used to include electron correlation.  Commonly used 
methods include density functional theory (DFT)~\cite{jones1989density}, configuration interaction 
(CI)~\cite{harrison1983full}, many body perturbation theory(MBPT)~\cite{moller1934note,handy1985convergence,
pendergast1978partial} and coupled cluster(CC)~\cite{bartlett2007coupled,coester1958bound,coester1960short,
vcivzek1966correlation,vcivzek1971correlation,paldus1972correlation,purvis1982full,raghavachari1989fifth}.  
The CC method has been regarded as the ``gold'' standard~\cite{bartlett2007coupled}.  These approaches, while 
very useful, have well-known deficiencies such as the approximate inclusion of correlation (DFT), size 
inconsistency (truncated CI, such as with single and double excitations or with single, double, and triple 
excitations), or non-variational energies (CC).  Therefore it is important to investigate alternative approaches.

There are three major numerical methods used to study strongly correlated many body systems. These are exact 
diagonalization, density matrix renormalization group (DMRG)~\cite{white1993density}, and QMC. Exact 
diagonalization is only feasible for small systems since it scales exponentially with the system size.
DMRG has become useful for certain classes of molecules with an  order 
of 50 strongly correlated electrons~\cite{chan2011density,marti2010density,wouters2014density}.

The Monte Carlo method was first introduced and developed by Fermi, Teller, and Metropolis~\cite{metropolis1953equation,
metropolis1949monte,fermi1948note}.  QMC, unlike exact diagonalization and DMRG, is a scalable method that 
can be applied to multi-dimensional lattice systems. However, QMC does have the minus sign problem in 
fermionic and frustrated quantum systems.

A variety of methods have been proposed to alleviate the minus sign problem in QMC.  These  include  
auxiliary field Monte Carlo~\cite{rom1998shifted}, 
shifted contour auxiliary field Monte Carlo~\cite{baer1998shifted},  and fixed node diffusion Monte Carlo~\cite{ceperley1980ground,
foulkes2001quantum}. More recently, a resummation path integral approach~\cite{thom2005combinatorial,
thom2008electron}, which is similar to the SiLK method,  phaseless 
auxiliary-field QMC~\cite{suewattana2007phaseless},  
a finite temperature version of diffusion Monte Carlo~\cite{brown2013path,brown2013exchange}, and full 
configuration interaction QMC~\cite{booth2009fermion,booth2013towards} have been developed.

The Sign-Learning Kink (SiLK) QMC algorithm  originally developed by Hall~\cite {hall2002adaptive,hall2002kink} 
can be used to overcome the minus sign problem.  
SiLK has previously been used to study the 3$\times$3 Hubbard 
model and atoms using a small basis set~\cite {hall2002adaptive,hall2002kink}.  An approximate version of the 
method has been used to study small molecules~\cite{hall2005simulation}.  This method uses a novel learning 
process to overcome the minus sign problem. 

The goal of this work is to investigate the ability of the SiLK method to reduce the sign problem and 
accurately calculate potential energy surfaces in model systems with relatively small basis sets. 
Investigation of the scalability of the method is left for future work.  Therefore, SiLK QMC calculations 
are performed on H$_{2}$O, N$_{2}$, and F$_{2}$ at a number of different geometries.   The results of 
the calculations are  compared to the results to exact diagonalization and a variety of quantum chemistry 
methods  and demonstrate that the SiLK method is accurate and that it reduces the minus sign problem for 
all geometries.

\section{SiLK Formalism and Algorithm}
\subsection{SiLK Formalism}
Assume there are a finite set of states composed of Slater determinants $\left\{ \alpha_{i}\right\}$ formed 
from orthogonal, one electron spin orbitals.  With Hamiltonian, $H$, and  $\beta = 1/k_{B}T$, the canonical partition function $Q$ can be written as
\begin{equation}
Q=Tr\left\{ e^{-\beta H}\right\} =\sum_{j}\left\langle \alpha _{j}|e^{-\beta
H}|\alpha _{j}\right\rangle.
\end{equation}
Using 
\begin{equation}
 e^{-\beta H}={(e^{-\beta H/P})}^P,
\end{equation}
and the identity
\begin{eqnarray}
 1 = \sum_{j_{i}} |\alpha_{j_{i}} \rangle \langle \alpha_{j_{i}}|\,,
\end{eqnarray}
the partition function becomes

\begin{eqnarray}
Q&=&\sum_{j_{1},j_{2},...,j_{P}}
\Braket{\alpha_{j_{1}}|\exp\left(-\frac{\beta}{P}H\right)|\alpha_{j_{2}}}
\Braket{\alpha_{j_{2}}|\exp\left(-\frac{\beta}{P}H\right)|\alpha_{j_{3}}}
\cdots \nonumber \\
&&
\Braket{\alpha_{j_{P}}|\exp\left(-\frac{\beta}{P}H\right)|\alpha_{j_{1}}} \, .
 \label{trotter}
\end{eqnarray}

$P$ is introduced as a discretization variable that allows for the evaluation of the matrix elements by expanding the exponential, \textit{vide infra}.
For a given set of $\{\alpha_{j_{i}}\}$, some of the matrix elements in Eqn.~\ref{trotter} may be diagonal. Thus, terms 
appearing in the summand may be classified by the number of off-diagonal matrix elements.  
In the SiLK formalism, off-diagonal matrix elements are referred to as kinks. By analytically 
summing over the diagonal matrix elements in Eqn.~\ref{trotter}, we obtain a kink-based version of the partition function. 
Defining
\begin{eqnarray}
x_{j}&\equiv&
\Braket{\alpha_{j}|\exp\left(-\frac{\beta}{P}H\right)|\alpha_{j}}\approx
\Braket{\alpha_{j}|         1-\frac{\beta}{P}H       |\alpha_{j}}+
  \mathscr{O}\left(\frac{1}{P^{2}}\right)\nonumber\\
&\approx&\exp\left(-\frac{\beta}{P}\Braket{\alpha_{j}|H|\alpha_{j}}\right)+\mathscr{O}\left(\frac{1}{P^{2}}\right),\\
t_{ij}&\equiv&
\Braket{\alpha_{i}|\exp\left(-\frac{\beta}{P}H\right)|\alpha_{j}}\approx
\Braket{\alpha_{i}|         1-\frac{\beta}{P}H       |\alpha_{j}}+
  \mathscr{O}\left(\frac{1}{P^{2}}\right)\nonumber\\
&\approx&\exp\left(-\frac{\beta}{P}\Braket{\alpha_{i}|H|\alpha_{j}}\right)-1+\mathscr{O}\left(\frac{1}{P^{2}}\right),  {\rm when} \, \, i\ne j,
\label{eqn:kink_def}
\end{eqnarray}
the result of this analytical summation is~\cite{hall2002adaptive}:
\begin{eqnarray}
Q&=&\lim_{P\rightarrow\infty}Q\left( P\right),\\
Q\left( P\right)  &=&\sum_{j}x_{j}^{P}+ \sum_{n=2}^{P}\frac{P}{n}\left(
\prod_{i=1}^{n}\sum_{j_{i}}\right) \left(
\prod_{k=1}^{n}t_{j_{k},j_{k+1}}\right) S\left( \left\{
x_{j}\right\} ,n,m,\left\{ s_{j}\right\} \right),
\label{equ:finaleqn}
\end{eqnarray}
where $n$ is the number of kinks, $m$ is the number of distinct $\alpha_{j}$'s in a given set of states with $n$ kinks, $s_{j}$ is the number of times the state $\alpha_{j}$ appears in a given set of states  with $n$ kinks, and 

\begin{eqnarray}
S\left( \left\{ x_{j}\right\} ,n,m,\left\{ s_{j}\right\} \right)
=\prod_{j=1}^{m}\left[ \frac{1}{\left( s_{j}-1\right) !}\frac{%
d^{s_{j}-1}}{dx_{j}^{s_{j}-1}}x_{j}^{s_{j}-1}\right]
\sum_{l=1}^{m}\frac{x_{l}^{P-n+m-1}}{\prod\limits_{i\neq l}\left(
x_{l}-x_{i}\right) }  \label{Final_S},
\end{eqnarray}
where $S$ may be evaluated recursively.  Due to the derivatives in Eqn.~\ref{Final_S}, it is possible for $S$ to be negative.  In addition, the off-diagonal matrix elements $t_{j_{k},j_{k+1}}$ can also be negative. 

\begin{figure}[ht]
\centerline{ \includegraphics[clip,scale=0.5]{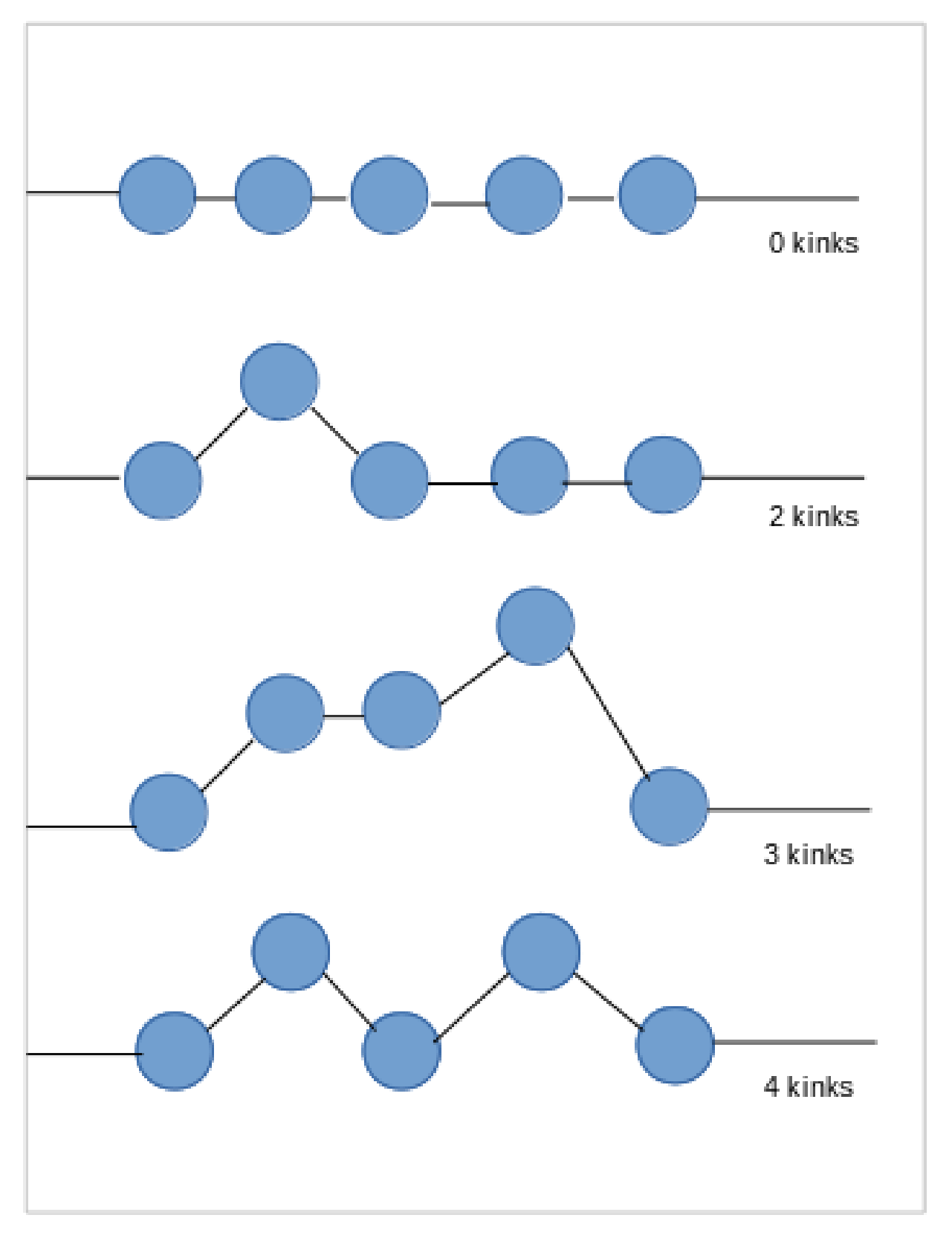}} 
\caption{Examples of different kink configurations that occur in Eqn.~\ref{trotter} when P = 5. Horizontal lines correspond to diagonal matrix elements and slanted lines correspond to off-diagonal matrix elements that are referred to as kinks.  The  lines at the beginning and end of a kink configuration wrap around due to the $Trace$ operation required by the partition function. 
 The zero kink configuration contains matrix elements for a single state, the two kink configuration 
contains matrix elements for just two states, etc.  Note that the number of kinks and the number of states are not necessarily equal to each other as it is seen in the four kink configuration.}
\label{fig:kink_definition} %
\end{figure}

Fig.~\ref{fig:kink_definition} depicts the types of kink configurations that appear in this sum over 
states. The top figure without kinks corresponds to the case where only diagonal matrix elements occur such
that $j_{1}=j_{2},\cdots$.  The second case contains two ``kinks'' where two identical off-diagonal 
matrix elements are introduced.  This so-called kink expansion was used in condensed matter physics by Anderson~\cite{anderson1970exact}
and later in the chemical physics literature by Wolynes~\cite{chiles1984monte}.

The first term is non-negative.  The $n=2$ is also non-negative since the off-diagonal matrix elements appear as $|t_{j_{1},j_{2}}|^{2}$  and with $s_{1}$ and $s_{2}$ = 1, $S > 0$.  Therefore the sign problem is due to terms with  $n\ge 3$.  The SiLK method uses a learning algorithm to construct new states as linear combinations of the initial $\{\alpha_{i}\} $ states that minimize the magnitude of the contributions from terms with $n\ge 3$ and thereby reduces or eliminates the sign problem.  

Eqn.~\ref{equ:finaleqn} has the form of a grand canonical partition function and thus Monte Carlo methods may be used to evaluate the partition function and its properties. Writing this equation as
\begin{eqnarray}
Q(P)&=&\sum_{n=0}^{P}\sum_{\{\alpha_{i}\}} \rho(n,\{\alpha_{i}\})
\end{eqnarray}
a Monte Carlo simulation will involve sampling different states and inserting and removing kinks.  The average energy of the system can be evaluated using $<E>=-\frac{d}{d\beta}\ln Q$,
\begin{eqnarray}
      <E>&=&-\frac{\sum_{n=0}^{P}\sum_{\{\alpha_{i}\}} \frac{d}{d\beta}\rho(n,\{\alpha_{i}\})}{\sum_{n=0}^{P}\sum_{\{\alpha_{i}\}} \rho(n,\{\alpha_{i}\})}\nonumber\\
&=&-\frac{\sum_{n=0}^{P}\sum_{\{\alpha_{i}\}}\frac{|\rho(n,\{\alpha_{i}\})|}{|\rho(n,\{\alpha_{i}\})|} \frac{d}{d\beta}\rho(n,\{\alpha_{i}\})}{\sum_{n=0}^{P}\sum_{\{\alpha_{i}\}}\frac{|\rho(n,\{\alpha_{i}\})|}{|\rho(n,\{\alpha_{i}\})|} \rho(n,\{\alpha_{i}\})}\nonumber\\
&=&-\frac{<\frac{1}{|\rho|}\frac{d}{d\beta}\rho>_{|\rho|}}{<sign(\rho)>_{|\rho|}}
\end{eqnarray}
 where kink configurations are sampled from $|\rho|$.  
 
\subsection{SiLK algorithm}
Simulations are performed in two stages.  The first stage is a ``learning'' period and is used to construct 
an improved description of the states of the system. We choose the lowest energy Hartree-Fock state as the 
initial state.  As the grand canonical simulation proceeds, additional states are inserted 
and removed and a list is maintained of states that have appeared.  At fixed intervals (30 iterations in our 
calculations) or when the number of kinks present at the end of a Monte Carlo pass exceeds a specified number 
(9 in our case),
the Hamiltonian 
is diagonalized in the sub-space of the states that have appeared since the last diagonalization (or the start of 
the simulation).  The state  is then set to the lowest energy state  (a zero-kink configuration) and the 
simulation is continued.  The learning period ends when there are only zero and two kinks configurations present 
for an extended number of Monte Carlo passes.  At this point, the expectation is that the partition function will 
be dominated by kink configurations with a small number of kinks (dominated
by configurations with 0 or 2 kinks) as the current set of states will better approximate the ground state of the 
system than the initial ones.  As currently implemented, the learning stage can be thought of as using 
the simulation to construct configuration interaction (CI) states.  In the present work, the learning period ranged
from 8,000 to 119,000 passes.  

The second stage in the simulation is the data acquisition during which the 
states are not modified and the grand canonical simulation proceeds in the standard way.  If the number of kinks 
increases dramatically during this stage (perhaps due to the simulation exploring a previously unexplored region 
of phase space) a diagonalization is performed and the second stage is restarted.  In the calculations where 
additional  diagonalizations were performed the diagonalization made an insignificant change in the ground and excited state energies.  Between 1,000 and 2,000 Monte Carlo passes were used in the second stage.  

The Monte Carlo algorithm consists of two types of moves: change of state and insertion/removal of states. The 
former is performed in the standard way using the Metropolis algorithm.  The latter uses the Metropolis algorithm 
as follows.  A potential new kink configuration $c'$ is sampled based on the current kink configuration $c$ using the 
normalized conditional probability $T(c'|c)$ and accepted with probability 
$A(c'|c) = min\left[1,\frac{\rho(c')T(c|c')}{\rho(c)T(c'|c)}\right]$.
If there are $n$ states  in the current kink configuration, there 
are $n+1$ places to insert a new state into the kink configuration (as state 1, state 2, ..., state $n+1$). There 
are $n$ ways to remove a state. We set 
\begin{eqnarray}
T(c'|c)&=& T_{remove}(c'|c) + T_{add}(c'|c) \label{normalizedprob}\label{eqn:sample1}\,,
\end{eqnarray}
with the probability of removing the state at location $k$ in the list of states
\begin{eqnarray}
T_{remove}(c'=\{n-1, k\}| c)&=&\frac{|\rho(1,2,...,k-1,k+1,...,n-1)|}{D(c'|c)}\label{remove}\,,
\end{eqnarray}
and with the probability of adding  $\alpha_j $ at location $k$:
\begin{eqnarray}
T_{add}(c'=\{n+1,k,\alpha_j \}| c)&=&\frac{|\rho(1,2,...,k-1,\alpha_j , k,k+1,...,n+1)|}{D(c'|c)}\label{add}\,,
\end{eqnarray}
with $D(c'|c)$ the normalization for the probability:
\begin{eqnarray}
D(c'|c) &=& \sum_k |\rho(1,2,...,k-1,k+1,...,n-1)| +\nonumber \\
&& \sum_k \sum_{\alpha_j } |\rho(1,2,...,k-1,\alpha_j , k,k+1,...n+1)|\,.
\end{eqnarray}
The acceptance probability is then
\begin{eqnarray}
A(c'|c)&=&\min\left[1,\frac{D(c'|c)}{D(c|c')}\right] \,.
\end{eqnarray}

\section{Results and Discussion}
  
We use the SiLK algorithm to calculate the energies of H$_{2}$O, N$_{2}$, and F$_{2}$ at selected bond lengths and 
bond angles for H$_{2}$O. The Cartesian Gaussian DZ basis 
set~\cite{feller1996role,schuchardt2007basis,dunning1970gaussian,dunning1977methods} is used in all calculations.
Computer memory constraints imposed by the current SiLK implementation 
dictates the number of determinants that can be used  in the calculations.
  The reason is that at present the CI coefficients 
for the ground and excited states must be stored.  Future work will focus on alleviating the memory issues. We therefore use either the full vector space of determinants generated by all possible excitations of the
Hartree-Fock determinant (Full Configuration Interaction, FCI) or the restricted vector spaces generated by the 
HF determinant and either all possible single and double excitations (SD) or all possible single, double and triple 
excitations (SDT) of the HF determinant.  In all cases, a comparison of the SiLK method to the exact result within the vector
space is made to assess, as mentioned in the introduction, the ability of SiLK to provide accurate results and alleviate the sign problem.
At each geometry, a Hartree-Fock computation using the NWChem \textit{ab initio} 
package~\cite{valiev2010nwchem} is used to generate the initial molecular orbitals from which the determinants 
are created.  Determinants corresponding to excited states are generated by excitations of all molecular 
orbitals except the core orbitals (the frozen core approximation).  Symmetry is used to restrict the determinants 
to those with the same symmetry as the ground Hartree-Fock state. 
For the calculations presented here, we use $C_{2v}$ spatial symmetry for H$_{2}$O and $D_{2h}$ spatial symmetry 
for N$_{2}$ and F$_{2}$ respectively.  We use $T=1$ K ($\beta=3\times10^{5}$ Hartree$^{-1}$).
Exact energies are obtained by numerical diagonalization for the SD and SDT vector spaces.  A series of calculations with increasing
values for $P$ were performed until a convergence in the energy was obtained.  The values of $P$ chosen for the reporting of
data ranged from $2\times10^{7}$ to $2\times10^{10}$.
The FCI calculations 
for H$_2$O is performed using Molpro~\cite{werner2012molpro,molpro}.
   
\begin{figure}[h!]
\centerline{ \includegraphics[clip,scale=0.42]{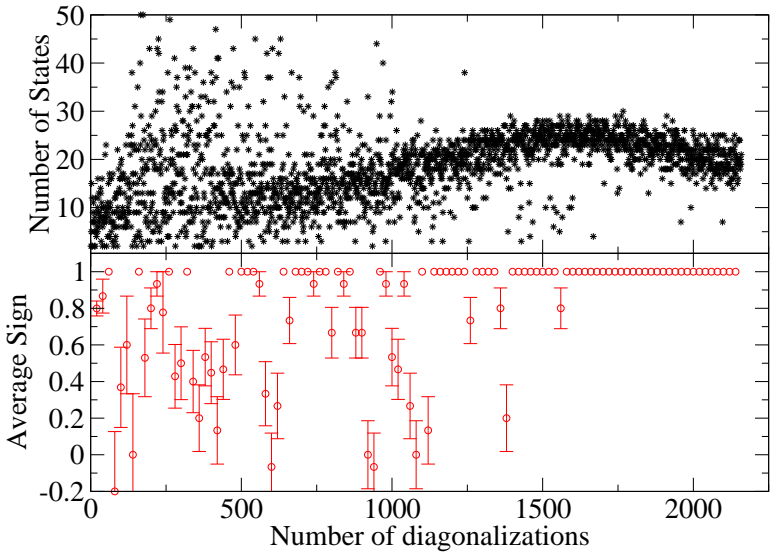}} 
\caption{The evolution of the sign during the SiLK learning period for H$_{2}$O using the DZ basis set at the FCI 
minimum energy geometry, $P=2\times10^{8}$.  The O-H bond length  is 1.84345 Bohr and the HOH angle is 110.565 degrees. The upper plot 
shows the number of states involved in each diagonalization, which was constrained to be less than or equal to 50.  
The lower plot shows the average sign evolution, averaged over every 20 diagonalizations, during the learning process.}
\label{fig:H2O_FCI_sign_evolution_states} %
\end{figure}

The ability of SiLK to address the sign problem is evaluated by following the evolution of the sign (for
clarity averaged over every 20 Monte Carlo steps) during the course of the learning period. Representative of the 
results from the different molecules
is the average sign for water at the minimum energy FCI geometry~\cite{saxe1981exact}. In this calculation, the 
maximum number of states included in a diagonalization is limited to 50 (the entire vector space had a dimension 
of 128,829).  The coarse-grained sign is shown in Fig.~\ref{fig:H2O_FCI_sign_evolution_states}.  
The sign fluctuates significantly for roughly the first 1,500 diagonalizations, but after approximately 1,600 
diagonalizations it remains 1.0.  In the upper panel of Fig.~\ref{fig:H2O_FCI_sign_evolution_states} the number of 
states involved in each diagonalization is shown.  After 1,600 diagonalizations, the coarse-grained sign remains 
at 1.0 even though the number of states involved in subsequent diagonalizations is approximately 20.
This indicates that kinks are being introduced during the Monte Carlo process but these are not affecting the sign.  
The  number of kinks averaged over the kink configurations  between 2 successive diagonalizations ranged from roughly 5
at the beginning of the learning period to roughly 2.5 at the end of the learning period.
An examination of the kink configurations after the learning period found that the configurations contain either 
zero and two kinks and therefore the average sign is 1.0.

Accurate calculations of potential energy surfaces are important in understanding reaction energetics and rates. 
The ability of SiLK and other quantum chemical methods to calculate potential energy surfaces is assessed 
for H$_2 $O, F$_{2}$, and N$_{2}$.  The goal of a successful method is to achieve the accuracy required to describe 
energetic differences encountered in chemical processes such as bond-breaking/bond-forming reactions and reaction 
activation energies.  This so-called ``chemical accuracy'' is approximately 0.1 -- 1 kcal/mol 
$\approx 10^{-3} - 10^{-4}$ Hartrees/mol~\cite{bash1996progress}.  

Several versions of truncated CC are used in this work.  The CCSD method uses  single and double excitations~\cite{purvis1982full}. The CCSDT
method uses   single, double, and triple excitations~\cite{noga1987full}. The CCSD(T) method uses  single, double, and
non-iterative inclusion of perturbative triples~\cite{raghavachari1989fifth} and is considered to be
the ``gold standard'' of ab initio quantum chemistry.
The MRCCSD(T)(2,2) and MRCCSD(T)(4,4) methods are multi reference CC (MRCC) methods with
single, double, and non-iterative inclusion of perturbative triples~\cite{bhaskaran2012implementation}.  A (2,2) calculation uses 2 electrons and 2 orbitals (one occupied and one virtual) to generate the model space for the MRCC calculation and a (4,4) calculation uses 4 electrons and 4 orbitals (two occupied and two virtual) to generate the model space for the MRCC calculation. We also use second order many body perturbation theory (MBPT(2)).  
The NWChem software package is used to perform all standard 
\textit{ab initio} calculations~\cite{valiev2010nwchem}.

\subsection{Water}

\begin{figure}[h!]
\centerline{\includegraphics[clip,scale=0.42]{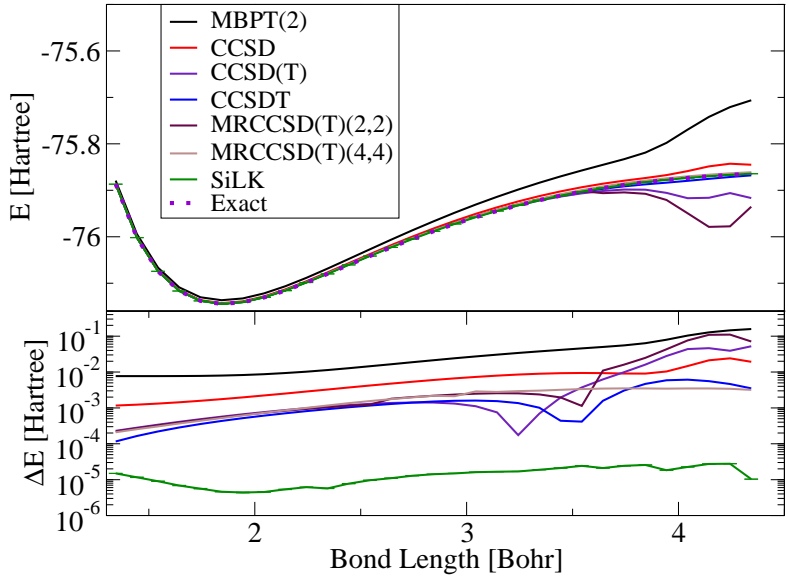}} 
\caption{Potential energy curve of  H$_2$O molecule as a function of OH bond length. A comparison of results 
obtained using MBPT(2), CCSD, CCSD(T), CCSDT, MRCCSD(T)(2,2),  MRCCSD(T)(4,4), and SiLK formalisms. The bottom 
plot displays the absolute error in the calculated energy for the different methods.
$P=2\times10^{8}$ is used for bond lengths in the range [1.34-3.64] Bohr and  $P=2\times10^{9}$ is used for bond 
lengths in the range [3.74-4.34] Bohr.}
\label{fig:H2O_FCI_frozen} %
\end{figure}

The H$_{2}$O molecule is used to assess the ability of SiLK to describe the variation of energy with bond length 
and bond angle in two separate calculations, one in which the bond length is varied and another in which the bond 
angle is varied. Fig.~\ref{fig:H2O_FCI_frozen} displays the energy and its absolute error as a function 
of bond length at fixed bond angle of 110.565 degrees as calculated by different methods. 
The SiLK method has an absolute error of $10^{-5}$ Hartree over the range of bond lengths studied, which is well 
below the desired chemical accuracy.  At the minimum energy geometry (bond length = 1.8434 Bohr), the exact 
energy is -76.14455299 Hartrees and the energy of the lowest energy SiLK state is -76.14454690 Hartrees, an error of 
$\approx 6 \times 10^{-6}$ Hartrees.  Therefore the SiLK procedure found an excellent approximation to the exact ground state.
SiLK is approximately one order of magnitude more accurate than the most 
accurate of the other methods in the comparison.  Notably, SiLK is accurate at the longer bond lengths (roughly 
two orders of magnitude more accurate than any other method) which is crucial to a description of bond 
dissociation and bond breaking processes. None of the other methods (except MRCCSD(T)(4,4)) achieves chemical 
accuracy over the entire range of bond lengths studied.

\begin{figure}[h!]
\centerline{\includegraphics[clip,scale=0.42]{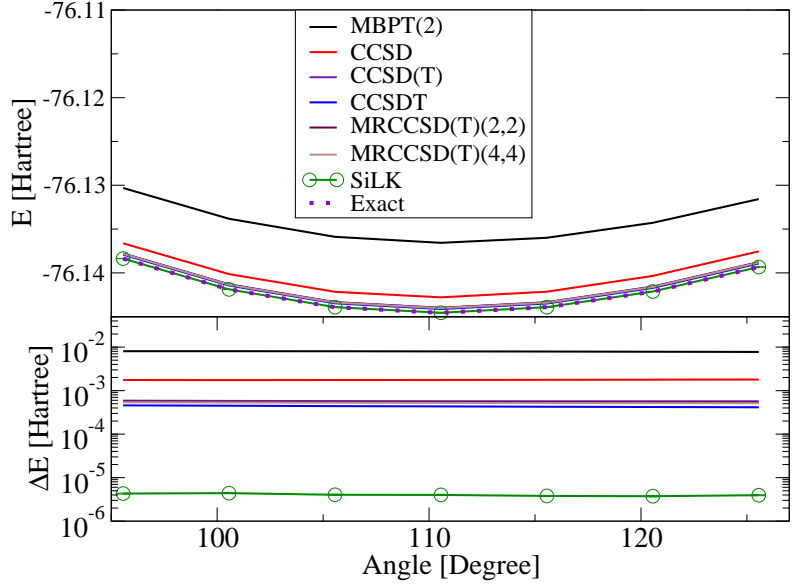}}
\caption{Potential energy curves of H$_2$O FCI vector space for the DZ basis~\cite{dunning1970gaussian}. 
A comparison of results obtained by SiLK with results from MBPT(2), CCSD, CCSD(T), CCSDT, MRCCSD(T)(2,2), and
MRCCSD(T)(4,4). Bottom plot displays the absolute error of 
energy. $P=2\times10^{8}$ is used for all angles in SiLK QMC.}
\label{fig:H2O_FCI_angle} %
\end{figure}

Then we use these methods to calculate the energy as a function of bond angle for the H$_{2}$O molecule with bond 
length 1.84345 Bohr. Fig.~\ref{fig:H2O_FCI_angle} displays the energies and absolute errors for bond angles ranging 
from 95 to 125 degrees. The SiLK method is approximately two orders of magnitude more accurate than the most accurate 
of the other methods. All methods except MBPT(2) and CCSD achieve chemical accuracy. 

\begin{figure}[h!]
\centerline{ \includegraphics[clip,scale=0.42]{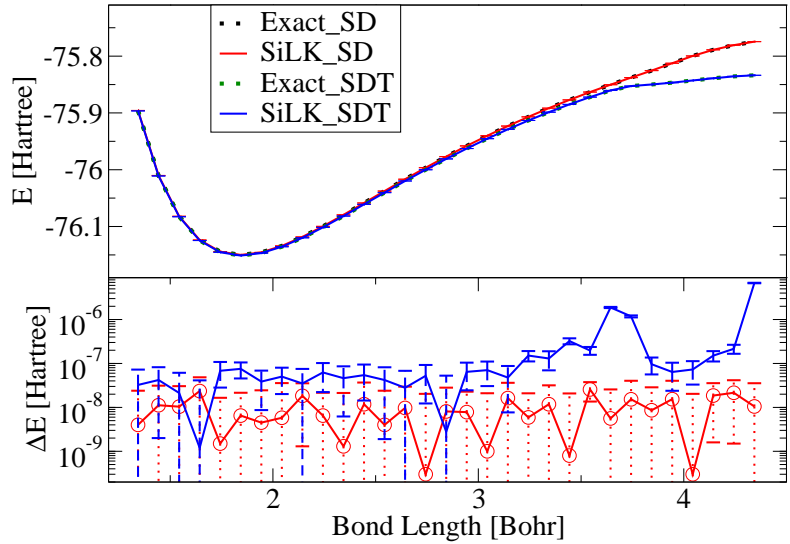}} 
\caption{Potential energy curves for H$_{2}$O molecule using the DZ basis and the SD and SDT vector spaces. The exact results obtained from exact
diagonalization and the SiLK results are shown. $P=2\times10^{10}$ is used for all the bond lengths in the SD and the SDT vector spaces.}
\label{fig:H2O_SDSDT} %
\end{figure}

\begin{figure}[h!]
\centerline{ \includegraphics[clip,scale=0.42]{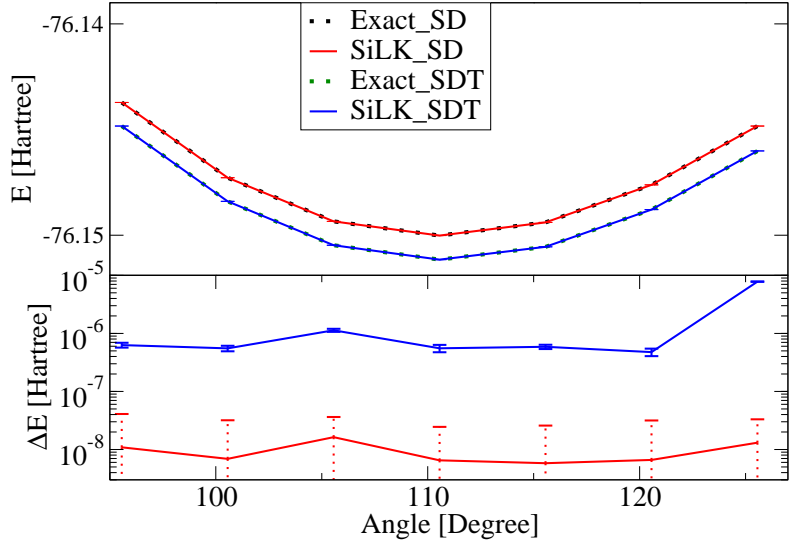}} 
\caption{Potential energy curves for the H$_{2}$O molecule within the SD and SDT vector spaces and the DZ basis. 
Exact results obtained from exact diagonalization and SiLK results are shown. $P=2\times10^{10}$ is used for all 
the angles in the SD vector space. Within the SDT space, $P=2\times10^{10}$ is used for angles in the range 
[95.565,120.565] and  $P=2\times10^{7}$ is used for the 125.565 degree calculation in SiLK QMC.}
\label{fig:H2O_SDSDT_angle} %
\end{figure}

It is also instructive to consider calculations restricted to just single and double excitations as sometimes 
computations based on such restricted vector spaces can yield useful results using significantly fewer computational 
resources.  Therefore, the SiLK algorithm is used to calculate the energies and absolute errors of the H$_{2}$O 
molecule as a function of bond length and bond angle.  Figs.~\ref{fig:H2O_SDSDT} and \ref{fig:H2O_SDSDT_angle} show 
their comparison with exact results.  The SiLK method is able to reproduce the exact results to $10^{-5}$ Hartree, 
well within chemical accuracy.

\subsection{Nitrogen}
\begin{figure}[h!]
\centerline{ \includegraphics[clip,scale=0.42]{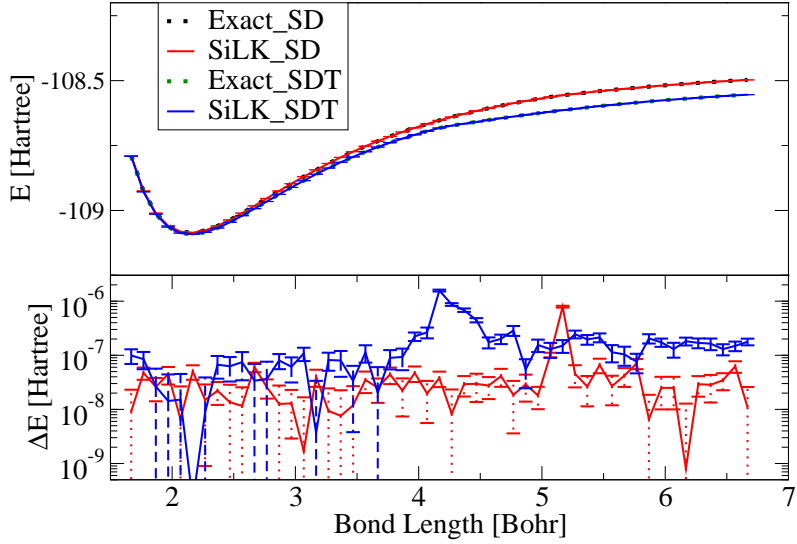}} 
\caption{Potential energy curves for N$_2$ molecule as a function of bond length using the SD and SDT vector spaces. 
The exact results obtained from exact
diagonalization and SiLK results are shown. $P=2\times10^{10}$ is used for all bond lengths.}
\label{fig:N2_SDSDT} 
\end{figure}

N$_{2}$ has a triple bond, which provides a challenging test for 
\textit{ab initio} methods due to its large electronic correlation~\cite{kowalski2000renormalized}. Due to memory 
limitations, the SiLK calculations were restricted to the Hartree-Fock determinant plus either the SD and SDT vector spaces.
 Fig.~\ref{fig:N2_SDSDT} shows that the SiLK QMC results converge to the exact result over  a wide range of bond 
lengths.  At the minimum energy geometry (bond length = 2.168 Bohr), the exact 
energy is -109.0858095 Hartrees and the energy of the lowest energy SiLK state is -109.0858094 Hartrees, an error of 
$\approx 1\times 10^{-7}$ Hartrees.  Therefore the SiLK procedure found an excellent approximation to the exact ground state.
The results demonstrate that the SiLK method
is suitable for multi reference systems such as N$_2$ where more than a single determinant is strongly coupled in 
the ground electronic state. 
\subsection{Fluoride}

\begin{figure}[h!]
\centerline{\includegraphics[clip,scale=0.42]{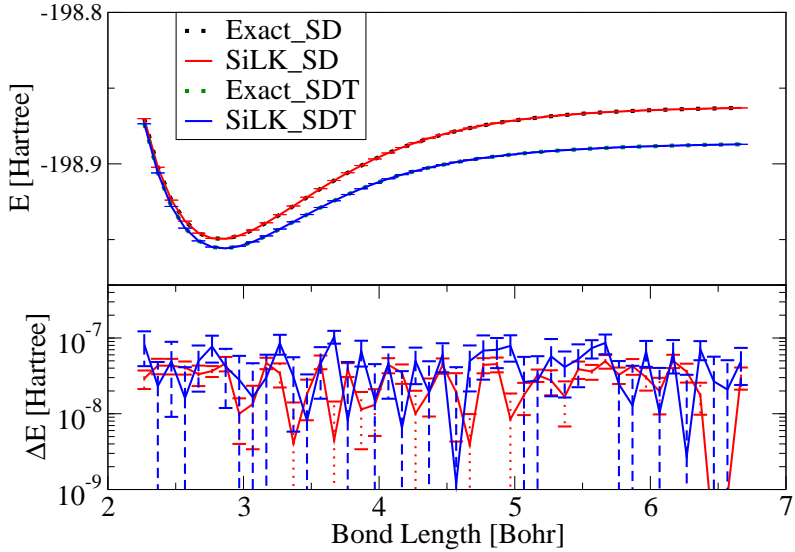}} 
\caption{Potential energy curves for F$_2$ molecule as a function of bond length and using the Hartree-Fock determinant plus either the SD and SDT vector spaces. The exact results obtained from exact
diagonalization and the SiLK results are shown. $P=2\times10^{10}$ is used.}
\label{fig:F2_SDSDT} %
\end{figure}

Electron correlations are  difficult to include in the simulation of F$_2$ as many determinants contribute small but important contributions to the total energy~\cite{kowalski2001comparison}.  This phenomenon is often referred to as dynamic correlation~\cite{mok1996dynamical}.
Therefore, the SiLK method is applied to the F$_{2}$ molecule.  As with Nitrogen, due to memory limitations, the SiLK calculations is limited to the  SD and SDT vector spaces.
 Fig.~\ref{fig:F2_SDSDT} shows that the SiLK QMC results converge to the exact results and demonstrate that SiLK is 
capable of accurately including dynamic correlation. At the minimum energy geometry (bond length = 2.86816 Bohr), the exact 
energy is -198.9494169 Hartrees and the energy of the lowest energy SiLK state is -198.9494169 Hartrees, an error of 
$< 1\times 10^{-7}$ Hartrees.  Therefore the SiLK procedure found an excellent approximation to the exact ground state.

\subsection{Scaling Analysis} 
\begin{figure}[h]
\centerline{ \includegraphics[clip,scale=0.42]{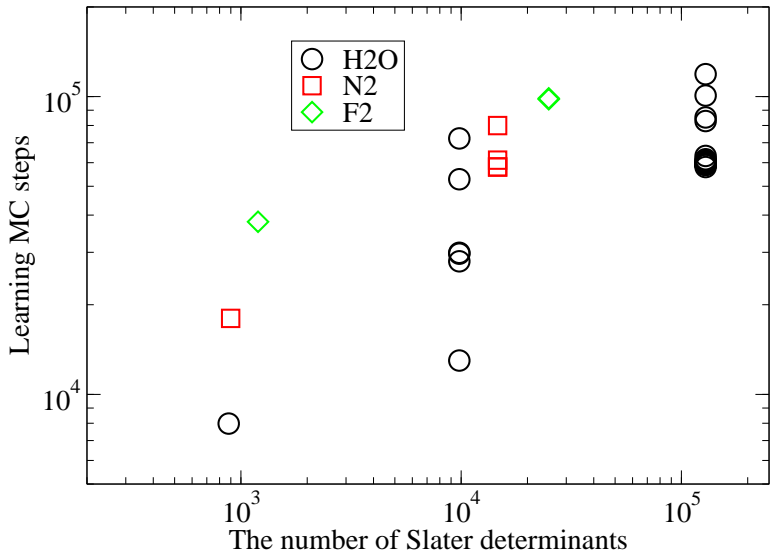}} 
\caption{The length of the learning period as a function of the number of Slater determinants.  The results for the SD, SDT, and FCI vector spaces for all molecules and geometries are shown.}
\label{fig:scaling_plot} %
\end{figure}

It is beyond the scope of this work to make a thorough analysis of the scaling of the SiLK method as the memory requirements of the current implementation of the SiLK method prohibits the use of a wide range of basis set size.  However, 
it is important to assess the scaling of the SiLK algorithm  with the size of the basis set and vector space.  No truncation methods, such as a truncation in the space of the single-particle density matrix \cite{maurits2014} will improve the efficiency of the algorithm for certain systems.  Therefore, a scaling analysis is presented for the current work using the relatively limited size of the vector spaces. If the SiLK algorithm  increases too quickly with the size of the vector space, the computational requirements for the SiLK method will make its use in the present form intractable. The dependence of the length of the learning period on the size of the vector space (number of determinants) is presented in Fig.~\ref{fig:scaling_plot}.  As expected, there is an increase in the size of the learning period.  However, a wider range of vector space sizes  is necessary to fully understand and quantify the scaling behavior. 

\section{Conclusions}
The minus sign problem in Quantum Monte Carlo simulations of frustrated or correlated electronic systems 
is a challenging problem.  It has even been suggested that a general solution of this problem is 
NP-complete~\cite{troyer2005computational}. Therefore, one should not expect an effective solution for 
all the Monte Carlo simulations which have  the minus sign problem. In this paper, we demonstrate that 
SiLK QMC can reduce the minus sign problem by using a learning stage that includes a diagonalization 
procedure. In this paper, we demonstrate that the energies obtained by the SiLK QMC match the  results 
from exact diagonalization and surpass the accuracy obtained using other quantum chemistry methods, 
particularly for geometries relatively far from equilibrium. In addition, SiLK can be applied to systems 
that require a multiple reference state approach.  An intriguing possibility for future work is to use  
the SiLK learning procedure in combination with other QMC algorithms to reduce the minus sign problem.

As the learning stage progresses, the states become more complicated linear
combinations of determinants so that more evaluations of matrix elements are required, thereby
increasing the computational expense.  However, at the same time the number of non-zero
matrix elements between these states decreases.  So, further optimization is possible by storing 
often-needed matrix elements in memory.  For example, storing the off-diagonal matrix elements 
between the ground and excited states yields a large speed up, since these are the only matrix 
elements required once the learning stage reaches the point where mostly zero and two kink 
configurations appear in the simulation. 
It is also possible to halt the learning stage at an 
earlier point, when the ground SiLK state is not as accurate an approximation to the exact ground 
state, but when the sign problem is alleviated but not eliminated and rely on the Monte Carlo 
sampling to provide the exact energy.  This would reduce the computational effort required to 
evaluate the matrix elements since fewer diagonalizations will have occurred.  An investigation 
of the efficacy of a shorter learning period is left for future work.

The SiLK method requires the knowledge of the off-diagonal matrix elements of the Hamiltonian.  
As the size of the system increases, the number of off-diagonal matrix elements increases 
factorially and it is not possible to store the matrix elements or the CI coefficients
for the ground and excited states that would allow for on-the-fly evaluation of the matrix elements.
As such, without a procedure to accurately truncate the number of determinants used to describe the 
ground and excited state wavefunctions, the use of the all possible determinants in a SiLK calculation  
will be limited to relatively small systems.  However, SiLK can certainly be used when determinants 
are restricted to, for example, single and double excitations.  Such truncated sets of determinants  
are often sufficient for the study of chemical systems.  In cases where restrictions to single
and double excitations are not sufficient, more sophisticated methods of truncation, such as the one developed
by Maurits\cite{maurits2004}, will be needed.

The SiLK QMC is a versatile method to calculate the ground state energy of molecular systems. Since 
the path integral formulation uses the canonical partition function it is possible to use the SiLK 
method to simulate the motion of the atoms at a finite temperature.  Future work will investigate 
the use of the SiLK method in finite temperature simulations.

\begin{acknowledgments}
This work is funded by the NSF EPSCoR LA-SiGMA project under award EPS-1003897 and the US Department of Energy 
under Contract DE-AC06.76RLO-1830. RWH acknowledges support from Dominican University of California's Lillian 
L.Y. Wang Yin, PhD Endowed Professorship of Chemistry. This work used the high performance computing resources 
provided by Louisiana State University (http://www.hpc.lsu.edu) and the computational resources of the Environmental 
Molecular Sciences Laboratory at  
Pacific Northwest National Laboratory (PNNL), which is sponsored by the Department of Energy Office of Biological 
and Environmental Research. 

\end{acknowledgments}


\begin{thebibliography}{55}%
\makeatletter
\providecommand \@ifxundefined [1]{%
 \@ifx{#1\undefined}
}%
\providecommand \@ifnum [1]{%
 \ifnum #1\expandafter \@firstoftwo
 \else \expandafter \@secondoftwo
 \fi
}%
\providecommand \@ifx [1]{%
 \ifx #1\expandafter \@firstoftwo
 \else \expandafter \@secondoftwo
 \fi
}%
\providecommand \natexlab [1]{#1}%
\providecommand \enquote  [1]{``#1''}%
\providecommand \bibnamefont  [1]{#1}%
\providecommand \bibfnamefont [1]{#1}%
\providecommand \citenamefont [1]{#1}%
\providecommand \href@noop [0]{\@secondoftwo}%
\providecommand \href [0]{\begingroup \@sanitize@url \@href}%
\providecommand \@href[1]{\@@startlink{#1}\@@href}%
\providecommand \@@href[1]{\endgroup#1\@@endlink}%
\providecommand \@sanitize@url [0]{\catcode `\\12\catcode `\$12\catcode
  `\&12\catcode `\#12\catcode `\^12\catcode `\_12\catcode `\%12\relax}%
\providecommand \@@startlink[1]{}%
\providecommand \@@endlink[0]{}%
\providecommand \url  [0]{\begingroup\@sanitize@url \@url }%
\providecommand \@url [1]{\endgroup\@href {#1}{\urlprefix }}%
\providecommand \urlprefix  [0]{URL }%
\providecommand \Eprint [0]{\href }%
\providecommand \doibase [0]{http://dx.doi.org/}%
\providecommand \selectlanguage [0]{\@gobble}%
\providecommand \bibinfo  [0]{\@secondoftwo}%
\providecommand \bibfield  [0]{\@secondoftwo}%
\providecommand \translation [1]{[#1]}%
\providecommand \BibitemOpen [0]{}%
\providecommand \bibitemStop [0]{}%
\providecommand \bibitemNoStop [0]{.\EOS\space}%
\providecommand \EOS [0]{\spacefactor3000\relax}%
\providecommand \BibitemShut  [1]{\csname bibitem#1\endcsname}%
\let\auto@bib@innerbib\@empty
\bibitem [{\citenamefont {Dreuw}\ and\ \citenamefont
  {Head-Gordon}(2005)}]{dreuw2005single}%
  \BibitemOpen
  \bibfield  {author} {\bibinfo {author} {\bibfnamefont {A.}~\bibnamefont
  {Dreuw}}\ and\ \bibinfo {author} {\bibfnamefont {M.}~\bibnamefont
  {Head-Gordon}},\ }\href@noop {} {\bibfield  {journal} {\bibinfo  {journal}
  {Chemical Reviews}\ }\textbf {\bibinfo {volume} {105}},\ \bibinfo {pages}
  {4009} (\bibinfo {year} {2005})}\BibitemShut {NoStop}%
\bibitem [{\citenamefont {Feynman}, \citenamefont {Hibbs},\ and\ \citenamefont
  {Styer}(2005)}]{feynman2005quantum}%
  \BibitemOpen
  \bibfield  {author} {\bibinfo {author} {\bibfnamefont {R.~P.}\ \bibnamefont
  {Feynman}}, \bibinfo {author} {\bibfnamefont {A.~R.}\ \bibnamefont {Hibbs}},
  \ and\ \bibinfo {author} {\bibfnamefont {D.~F.}\ \bibnamefont {Styer}},\
  }\href@noop {} {\emph {\bibinfo {title} {Quantum mechanics and path
  integrals: Emended edition}}}\ (\bibinfo  {publisher} {Dover Publications},\
  \bibinfo {year} {2005})\BibitemShut {NoStop}%
\bibitem [{\citenamefont {Foulkes}\ \emph {et~al.}(2001)\citenamefont
  {Foulkes}, \citenamefont {Mitas}, \citenamefont {Needs},\ and\ \citenamefont
  {Rajagopal}}]{foulkes2001quantum}%
  \BibitemOpen
  \bibfield  {author} {\bibinfo {author} {\bibfnamefont {W.}~\bibnamefont
  {Foulkes}}, \bibinfo {author} {\bibfnamefont {L.}~\bibnamefont {Mitas}},
  \bibinfo {author} {\bibfnamefont {R.}~\bibnamefont {Needs}}, \ and\ \bibinfo
  {author} {\bibfnamefont {G.}~\bibnamefont {Rajagopal}},\ }\href@noop {}
  {\bibfield  {journal} {\bibinfo  {journal} {Reviews of Modern Physics}\
  }\textbf {\bibinfo {volume} {73}},\ \bibinfo {pages} {33} (\bibinfo {year}
  {2001})}\BibitemShut {NoStop}%
\bibitem [{\citenamefont {Troyer}\ and\ \citenamefont
  {Wiese}(2005)}]{troyer2005computational}%
  \BibitemOpen
  \bibfield  {author} {\bibinfo {author} {\bibfnamefont {M.}~\bibnamefont
  {Troyer}}\ and\ \bibinfo {author} {\bibfnamefont {U.-J.}\ \bibnamefont
  {Wiese}},\ }\href@noop {} {\bibfield  {journal} {\bibinfo  {journal}
  {Physical Review Letters}\ }\textbf {\bibinfo {volume} {94}},\ \bibinfo
  {pages} {170201} (\bibinfo {year} {2005})}\BibitemShut {NoStop}%
\bibitem [{\citenamefont {Loh~Jr}\ \emph {et~al.}(1990)\citenamefont {Loh~Jr},
  \citenamefont {Gubernatis}, \citenamefont {Scalettar}, \citenamefont {White},
  \citenamefont {Scalapino},\ and\ \citenamefont {Sugar}}]{loh1990sign}%
  \BibitemOpen
  \bibfield  {author} {\bibinfo {author} {\bibfnamefont {E.}~\bibnamefont
  {Loh~Jr}}, \bibinfo {author} {\bibfnamefont {J.}~\bibnamefont {Gubernatis}},
  \bibinfo {author} {\bibfnamefont {R.}~\bibnamefont {Scalettar}}, \bibinfo
  {author} {\bibfnamefont {S.}~\bibnamefont {White}}, \bibinfo {author}
  {\bibfnamefont {D.}~\bibnamefont {Scalapino}}, \ and\ \bibinfo {author}
  {\bibfnamefont {R.}~\bibnamefont {Sugar}},\ }\href@noop {} {\bibfield
  {journal} {\bibinfo  {journal} {Physical Review B}\ }\textbf {\bibinfo
  {volume} {41}},\ \bibinfo {pages} {9301} (\bibinfo {year}
  {1990})}\BibitemShut {NoStop}%
\bibitem [{\citenamefont {Jones}\ and\ \citenamefont
  {Gunnarsson}(1989)}]{jones1989density}%
  \BibitemOpen
  \bibfield  {author} {\bibinfo {author} {\bibfnamefont {R.~O.}\ \bibnamefont
  {Jones}}\ and\ \bibinfo {author} {\bibfnamefont {O.}~\bibnamefont
  {Gunnarsson}},\ }\href@noop {} {\bibfield  {journal} {\bibinfo  {journal}
  {Reviews of Modern Physics}\ }\textbf {\bibinfo {volume} {61}},\ \bibinfo
  {pages} {689} (\bibinfo {year} {1989})}\BibitemShut {NoStop}%
\bibitem [{\citenamefont {Harrison}\ and\ \citenamefont
  {Handy}(1983)}]{harrison1983full}%
  \BibitemOpen
  \bibfield  {author} {\bibinfo {author} {\bibfnamefont {R.}~\bibnamefont
  {Harrison}}\ and\ \bibinfo {author} {\bibfnamefont {N.}~\bibnamefont
  {Handy}},\ }\href@noop {} {\bibfield  {journal} {\bibinfo  {journal}
  {Chemical Physics Letters}\ }\textbf {\bibinfo {volume} {95}},\ \bibinfo
  {pages} {386} (\bibinfo {year} {1983})}\BibitemShut {NoStop}%
\bibitem [{\citenamefont {M{\o}ller}\ and\ \citenamefont
  {Plesset}(1934)}]{moller1934note}%
  \BibitemOpen
  \bibfield  {author} {\bibinfo {author} {\bibfnamefont {C.}~\bibnamefont
  {M{\o}ller}}\ and\ \bibinfo {author} {\bibfnamefont {M.~S.}\ \bibnamefont
  {Plesset}},\ }\href@noop {} {\bibfield  {journal} {\bibinfo  {journal}
  {Physical Review}\ }\textbf {\bibinfo {volume} {46}},\ \bibinfo {pages} {618}
  (\bibinfo {year} {1934})}\BibitemShut {NoStop}%
\bibitem [{\citenamefont {Handy}, \citenamefont {Knowles},\ and\ \citenamefont
  {Somasundram}(1985)}]{handy1985convergence}%
  \BibitemOpen
  \bibfield  {author} {\bibinfo {author} {\bibfnamefont {N.}~\bibnamefont
  {Handy}}, \bibinfo {author} {\bibfnamefont {P.}~\bibnamefont {Knowles}}, \
  and\ \bibinfo {author} {\bibfnamefont {K.}~\bibnamefont {Somasundram}},\
  }\href@noop {} {\bibfield  {journal} {\bibinfo  {journal} {Theoretica Chimica
  Acta}\ }\textbf {\bibinfo {volume} {68}},\ \bibinfo {pages} {87} (\bibinfo
  {year} {1985})}\BibitemShut {NoStop}%
\bibitem [{\citenamefont {Pendergast}\ and\ \citenamefont
  {Hayes}(1978)}]{pendergast1978partial}%
  \BibitemOpen
  \bibfield  {author} {\bibinfo {author} {\bibfnamefont {P.}~\bibnamefont
  {Pendergast}}\ and\ \bibinfo {author} {\bibfnamefont {E.~F.}\ \bibnamefont
  {Hayes}},\ }\href@noop {} {\bibfield  {journal} {\bibinfo  {journal} {Journal
  of Computational Physics}\ }\textbf {\bibinfo {volume} {26}},\ \bibinfo
  {pages} {236} (\bibinfo {year} {1978})}\BibitemShut {NoStop}%
\bibitem [{\citenamefont {Bartlett}\ and\ \citenamefont
  {Musia{\l}}(2007)}]{bartlett2007coupled}%
  \BibitemOpen
  \bibfield  {author} {\bibinfo {author} {\bibfnamefont {R.~J.}\ \bibnamefont
  {Bartlett}}\ and\ \bibinfo {author} {\bibfnamefont {M.}~\bibnamefont
  {Musia{\l}}},\ }\href@noop {} {\bibfield  {journal} {\bibinfo  {journal}
  {Reviews of Modern Physics}\ }\textbf {\bibinfo {volume} {79}},\ \bibinfo
  {pages} {291} (\bibinfo {year} {2007})}\BibitemShut {NoStop}%
\bibitem [{\citenamefont {Coester}(1958)}]{coester1958bound}%
  \BibitemOpen
  \bibfield  {author} {\bibinfo {author} {\bibfnamefont {F.}~\bibnamefont
  {Coester}},\ }\href@noop {} {\bibfield  {journal} {\bibinfo  {journal}
  {Nuclear Physics}\ }\textbf {\bibinfo {volume} {7}},\ \bibinfo {pages} {421}
  (\bibinfo {year} {1958})}\BibitemShut {NoStop}%
\bibitem [{\citenamefont {Coester}\ and\ \citenamefont
  {K{\"u}mmel}(1960)}]{coester1960short}%
  \BibitemOpen
  \bibfield  {author} {\bibinfo {author} {\bibfnamefont {F.}~\bibnamefont
  {Coester}}\ and\ \bibinfo {author} {\bibfnamefont {H.}~\bibnamefont
  {K{\"u}mmel}},\ }\href@noop {} {\bibfield  {journal} {\bibinfo  {journal}
  {Nuclear Physics}\ }\textbf {\bibinfo {volume} {17}},\ \bibinfo {pages} {477}
  (\bibinfo {year} {1960})}\BibitemShut {NoStop}%
\bibitem [{\citenamefont
  {{\v{C}}{\'\i}{\v{z}}ek}(1966)}]{vcivzek1966correlation}%
  \BibitemOpen
  \bibfield  {author} {\bibinfo {author} {\bibfnamefont {J.}~\bibnamefont
  {{\v{C}}{\'\i}{\v{z}}ek}},\ }\href@noop {} {\bibfield  {journal} {\bibinfo
  {journal} {The Journal of Chemical Physics}\ }\textbf {\bibinfo {volume}
  {45}},\ \bibinfo {pages} {4256} (\bibinfo {year} {1966})}\BibitemShut
  {NoStop}%
\bibitem [{\citenamefont {{\v{C}}{\'\i}{\v{z}}ek}\ and\ \citenamefont
  {Paldus}(1971)}]{vcivzek1971correlation}%
  \BibitemOpen
  \bibfield  {author} {\bibinfo {author} {\bibfnamefont {J.}~\bibnamefont
  {{\v{C}}{\'\i}{\v{z}}ek}}\ and\ \bibinfo {author} {\bibfnamefont
  {J.}~\bibnamefont {Paldus}},\ }\href@noop {} {\bibfield  {journal} {\bibinfo
  {journal} {International Journal of Quantum Chemistry}\ }\textbf {\bibinfo
  {volume} {5}},\ \bibinfo {pages} {359} (\bibinfo {year} {1971})}\BibitemShut
  {NoStop}%
\bibitem [{\citenamefont {Paldus}, \citenamefont {{\v{C}}{\'\i}{\v{z}}ek},\
  and\ \citenamefont {Shavitt}(1972)}]{paldus1972correlation}%
  \BibitemOpen
  \bibfield  {author} {\bibinfo {author} {\bibfnamefont {J.}~\bibnamefont
  {Paldus}}, \bibinfo {author} {\bibfnamefont {J.}~\bibnamefont
  {{\v{C}}{\'\i}{\v{z}}ek}}, \ and\ \bibinfo {author} {\bibfnamefont
  {I.}~\bibnamefont {Shavitt}},\ }\href@noop {} {\bibfield  {journal} {\bibinfo
   {journal} {Physical Review A}\ }\textbf {\bibinfo {volume} {5}},\ \bibinfo
  {pages} {50} (\bibinfo {year} {1972})}\BibitemShut {NoStop}%
\bibitem [{\citenamefont {Purvis~III}\ and\ \citenamefont
  {Bartlett}(1982)}]{purvis1982full}%
  \BibitemOpen
  \bibfield  {author} {\bibinfo {author} {\bibfnamefont {G.~D.}\ \bibnamefont
  {Purvis~III}}\ and\ \bibinfo {author} {\bibfnamefont {R.~J.}\ \bibnamefont
  {Bartlett}},\ }\href@noop {} {\bibfield  {journal} {\bibinfo  {journal}
  {Journal of Chemical Physics}\ }\textbf {\bibinfo {volume} {76}},\ \bibinfo
  {pages} {1910} (\bibinfo {year} {1982})}\BibitemShut {NoStop}%
\bibitem [{\citenamefont {Raghavachari}\ \emph {et~al.}(1989)\citenamefont
  {Raghavachari}, \citenamefont {Trucks}, \citenamefont {Pople},\ and\
  \citenamefont {Head-Gordon}}]{raghavachari1989fifth}%
  \BibitemOpen
  \bibfield  {author} {\bibinfo {author} {\bibfnamefont {K.}~\bibnamefont
  {Raghavachari}}, \bibinfo {author} {\bibfnamefont {G.~W.}\ \bibnamefont
  {Trucks}}, \bibinfo {author} {\bibfnamefont {J.~A.}\ \bibnamefont {Pople}}, \
  and\ \bibinfo {author} {\bibfnamefont {M.}~\bibnamefont {Head-Gordon}},\
  }\href@noop {} {\bibfield  {journal} {\bibinfo  {journal} {Chemical Physics
  Letters}\ }\textbf {\bibinfo {volume} {157}},\ \bibinfo {pages} {479}
  (\bibinfo {year} {1989})}\BibitemShut {NoStop}%
\bibitem [{\citenamefont {White}(1993)}]{white1993density}%
  \BibitemOpen
  \bibfield  {author} {\bibinfo {author} {\bibfnamefont {S.~R.}\ \bibnamefont
  {White}},\ }\href@noop {} {\bibfield  {journal} {\bibinfo  {journal}
  {Physical Review B}\ }\textbf {\bibinfo {volume} {48}},\ \bibinfo {pages}
  {10345} (\bibinfo {year} {1993})}\BibitemShut {NoStop}%
\bibitem [{\citenamefont {Chan}\ and\ \citenamefont
  {Sharma}(2011)}]{chan2011density}%
  \BibitemOpen
  \bibfield  {author} {\bibinfo {author} {\bibfnamefont {G.~K.-L.}\
  \bibnamefont {Chan}}\ and\ \bibinfo {author} {\bibfnamefont {S.}~\bibnamefont
  {Sharma}},\ }\href@noop {} {\bibfield  {journal} {\bibinfo  {journal} {Annual
  Review of Physical Chemistry}\ }\textbf {\bibinfo {volume} {62}},\ \bibinfo
  {pages} {465} (\bibinfo {year} {2011})}\BibitemShut {NoStop}%
\bibitem [{\citenamefont {Marti}\ and\ \citenamefont
  {Reiher}(2010)}]{marti2010density}%
  \BibitemOpen
  \bibfield  {author} {\bibinfo {author} {\bibfnamefont {K.~H.}\ \bibnamefont
  {Marti}}\ and\ \bibinfo {author} {\bibfnamefont {M.}~\bibnamefont {Reiher}},\
  }\href@noop {} {\bibfield  {journal} {\bibinfo  {journal} {Zeitschrift
  f{\"u}r Physikalische Chemie International Journal of Research in Physical
  Chemistry and Chemical Physics}\ }\textbf {\bibinfo {volume} {224}},\
  \bibinfo {pages} {583} (\bibinfo {year} {2010})}\BibitemShut {NoStop}%
\bibitem [{\citenamefont {Wouters}\ and\ \citenamefont
  {Van~Neck}(2014)}]{wouters2014density}%
  \BibitemOpen
  \bibfield  {author} {\bibinfo {author} {\bibfnamefont {S.}~\bibnamefont
  {Wouters}}\ and\ \bibinfo {author} {\bibfnamefont {D.}~\bibnamefont
  {Van~Neck}},\ }\href@noop {} {\bibfield  {journal} {\bibinfo  {journal} {The
  European Physical Journal D}\ }\textbf {\bibinfo {volume} {68}},\ \bibinfo
  {pages} {1} (\bibinfo {year} {2014})}\BibitemShut {NoStop}%
\bibitem [{\citenamefont {Metropolis}\ \emph {et~al.}(1953)\citenamefont
  {Metropolis}, \citenamefont {Rosenbluth}, \citenamefont {Rosenbluth},
  \citenamefont {Teller},\ and\ \citenamefont
  {Teller}}]{metropolis1953equation}%
  \BibitemOpen
  \bibfield  {author} {\bibinfo {author} {\bibfnamefont {N.}~\bibnamefont
  {Metropolis}}, \bibinfo {author} {\bibfnamefont {A.~W.}\ \bibnamefont
  {Rosenbluth}}, \bibinfo {author} {\bibfnamefont {M.~N.}\ \bibnamefont
  {Rosenbluth}}, \bibinfo {author} {\bibfnamefont {A.~H.}\ \bibnamefont
  {Teller}}, \ and\ \bibinfo {author} {\bibfnamefont {E.}~\bibnamefont
  {Teller}},\ }\href@noop {} {\bibfield  {journal} {\bibinfo  {journal}
  {Journal of Chemical Physics}\ }\textbf {\bibinfo {volume} {21}},\ \bibinfo
  {pages} {1087} (\bibinfo {year} {1953})}\BibitemShut {NoStop}%
\bibitem [{\citenamefont {Metropolis}\ and\ \citenamefont
  {Ulam}(1949)}]{metropolis1949monte}%
  \BibitemOpen
  \bibfield  {author} {\bibinfo {author} {\bibfnamefont {N.}~\bibnamefont
  {Metropolis}}\ and\ \bibinfo {author} {\bibfnamefont {S.}~\bibnamefont
  {Ulam}},\ }\href@noop {} {\bibfield  {journal} {\bibinfo  {journal} {Journal
  of the American Statistical Association}\ }\textbf {\bibinfo {volume} {44}},\
  \bibinfo {pages} {335} (\bibinfo {year} {1949})}\BibitemShut {NoStop}%
\bibitem [{\citenamefont {Fermi}\ and\ \citenamefont
  {Richtmyer}(1948)}]{fermi1948note}%
  \BibitemOpen
  \bibfield  {author} {\bibinfo {author} {\bibfnamefont {E.}~\bibnamefont
  {Fermi}}\ and\ \bibinfo {author} {\bibfnamefont {R.}~\bibnamefont
  {Richtmyer}},\ }\href@noop {} {\enquote {\bibinfo {title} {{Note on
  census-taking in Monte-Carlo calculations}},}\ }\bibinfo {type} {Tech. Rep.}\
  (\bibinfo  {institution} {Los Alamos Scientific Lab.},\ \bibinfo {year}
  {1948})\BibitemShut {NoStop}%
\bibitem [{\citenamefont {Rom}\ \emph {et~al.}(1998)\citenamefont {Rom},
  \citenamefont {Fattal}, \citenamefont {Gupta}, \citenamefont {Carter},\ and\
  \citenamefont {Neuhauser}}]{rom1998shifted}%
  \BibitemOpen
  \bibfield  {author} {\bibinfo {author} {\bibfnamefont {N.}~\bibnamefont
  {Rom}}, \bibinfo {author} {\bibfnamefont {E.}~\bibnamefont {Fattal}},
  \bibinfo {author} {\bibfnamefont {A.~K.}\ \bibnamefont {Gupta}}, \bibinfo
  {author} {\bibfnamefont {E.~A.}\ \bibnamefont {Carter}}, \ and\ \bibinfo
  {author} {\bibfnamefont {D.}~\bibnamefont {Neuhauser}},\ }\href@noop {}
  {\bibfield  {journal} {\bibinfo  {journal} {Journal of Chemical Physics}\
  }\textbf {\bibinfo {volume} {109}},\ \bibinfo {pages} {8241} (\bibinfo {year}
  {1998})}\BibitemShut {NoStop}%
\bibitem [{\citenamefont {Baer}, \citenamefont {Head-Gordon},\ and\
  \citenamefont {Neuhauser}(1998)}]{baer1998shifted}%
  \BibitemOpen
  \bibfield  {author} {\bibinfo {author} {\bibfnamefont {R.}~\bibnamefont
  {Baer}}, \bibinfo {author} {\bibfnamefont {M.}~\bibnamefont {Head-Gordon}}, \
  and\ \bibinfo {author} {\bibfnamefont {D.}~\bibnamefont {Neuhauser}},\
  }\href@noop {} {\bibfield  {journal} {\bibinfo  {journal} {Journal of
  Chemical Physics}\ }\textbf {\bibinfo {volume} {109}},\ \bibinfo {pages}
  {6219} (\bibinfo {year} {1998})}\BibitemShut {NoStop}%
\bibitem [{\citenamefont {Ceperley}\ and\ \citenamefont
  {Alder}(1980)}]{ceperley1980ground}%
  \BibitemOpen
  \bibfield  {author} {\bibinfo {author} {\bibfnamefont {D.~M.}\ \bibnamefont
  {Ceperley}}\ and\ \bibinfo {author} {\bibfnamefont {B.}~\bibnamefont
  {Alder}},\ }\href@noop {} {\bibfield  {journal} {\bibinfo  {journal}
  {Physical Review Letters}\ }\textbf {\bibinfo {volume} {45}},\ \bibinfo
  {pages} {566} (\bibinfo {year} {1980})}\BibitemShut {NoStop}%
\bibitem [{\citenamefont {Thom}\ and\ \citenamefont
  {Alavi}(2005)}]{thom2005combinatorial}%
  \BibitemOpen
  \bibfield  {author} {\bibinfo {author} {\bibfnamefont {A.~J.}\ \bibnamefont
  {Thom}}\ and\ \bibinfo {author} {\bibfnamefont {A.}~\bibnamefont {Alavi}},\
  }\href@noop {} {\bibfield  {journal} {\bibinfo  {journal} {The Journal of
  chemical physics}\ }\textbf {\bibinfo {volume} {123}},\ \bibinfo {pages}
  {204106} (\bibinfo {year} {2005})}\BibitemShut {NoStop}%
\bibitem [{\citenamefont {Thom}, \citenamefont {Booth},\ and\ \citenamefont
  {Alavi}(2008)}]{thom2008electron}%
  \BibitemOpen
  \bibfield  {author} {\bibinfo {author} {\bibfnamefont {A.~J.}\ \bibnamefont
  {Thom}}, \bibinfo {author} {\bibfnamefont {G.~H.}\ \bibnamefont {Booth}}, \
  and\ \bibinfo {author} {\bibfnamefont {A.}~\bibnamefont {Alavi}},\
  }\href@noop {} {\bibfield  {journal} {\bibinfo  {journal} {Physical Chemistry
  Chemical Physics}\ }\textbf {\bibinfo {volume} {10}},\ \bibinfo {pages} {652}
  (\bibinfo {year} {2008})}\BibitemShut {NoStop}%
\bibitem [{\citenamefont {Suewattana}\ \emph {et~al.}(2007)\citenamefont
  {Suewattana}, \citenamefont {Purwanto}, \citenamefont {Zhang}, \citenamefont
  {Krakauer},\ and\ \citenamefont {Walter}}]{suewattana2007phaseless}%
  \BibitemOpen
  \bibfield  {author} {\bibinfo {author} {\bibfnamefont {M.}~\bibnamefont
  {Suewattana}}, \bibinfo {author} {\bibfnamefont {W.}~\bibnamefont
  {Purwanto}}, \bibinfo {author} {\bibfnamefont {S.}~\bibnamefont {Zhang}},
  \bibinfo {author} {\bibfnamefont {H.}~\bibnamefont {Krakauer}}, \ and\
  \bibinfo {author} {\bibfnamefont {E.~J.}\ \bibnamefont {Walter}},\
  }\href@noop {} {\bibfield  {journal} {\bibinfo  {journal} {Physical Review
  B}\ }\textbf {\bibinfo {volume} {75}},\ \bibinfo {pages} {245123} (\bibinfo
  {year} {2007})}\BibitemShut {NoStop}%
\bibitem [{\citenamefont {Brown}\ \emph
  {et~al.}(2013{\natexlab{a}})\citenamefont {Brown}, \citenamefont {Clark},
  \citenamefont {DuBois},\ and\ \citenamefont {Ceperley}}]{brown2013path}%
  \BibitemOpen
  \bibfield  {author} {\bibinfo {author} {\bibfnamefont {E.~W.}\ \bibnamefont
  {Brown}}, \bibinfo {author} {\bibfnamefont {B.~K.}\ \bibnamefont {Clark}},
  \bibinfo {author} {\bibfnamefont {J.~L.}\ \bibnamefont {DuBois}}, \ and\
  \bibinfo {author} {\bibfnamefont {D.~M.}\ \bibnamefont {Ceperley}},\
  }\href@noop {} {\bibfield  {journal} {\bibinfo  {journal} {Physical review
  letters}\ }\textbf {\bibinfo {volume} {110}},\ \bibinfo {pages} {146405}
  (\bibinfo {year} {2013}{\natexlab{a}})}\BibitemShut {NoStop}%
\bibitem [{\citenamefont {Brown}\ \emph
  {et~al.}(2013{\natexlab{b}})\citenamefont {Brown}, \citenamefont {DuBois},
  \citenamefont {Holzmann},\ and\ \citenamefont
  {Ceperley}}]{brown2013exchange}%
  \BibitemOpen
  \bibfield  {author} {\bibinfo {author} {\bibfnamefont {E.~W.}\ \bibnamefont
  {Brown}}, \bibinfo {author} {\bibfnamefont {J.~L.}\ \bibnamefont {DuBois}},
  \bibinfo {author} {\bibfnamefont {M.}~\bibnamefont {Holzmann}}, \ and\
  \bibinfo {author} {\bibfnamefont {D.~M.}\ \bibnamefont {Ceperley}},\
  }\href@noop {} {\bibfield  {journal} {\bibinfo  {journal} {Physical Review
  B}\ }\textbf {\bibinfo {volume} {88}},\ \bibinfo {pages} {081102} (\bibinfo
  {year} {2013}{\natexlab{b}})}\BibitemShut {NoStop}%
\bibitem [{\citenamefont {Booth}, \citenamefont {Thom},\ and\ \citenamefont
  {Alavi}(2009)}]{booth2009fermion}%
  \BibitemOpen
  \bibfield  {author} {\bibinfo {author} {\bibfnamefont {G.~H.}\ \bibnamefont
  {Booth}}, \bibinfo {author} {\bibfnamefont {A.~J.}\ \bibnamefont {Thom}}, \
  and\ \bibinfo {author} {\bibfnamefont {A.}~\bibnamefont {Alavi}},\
  }\href@noop {} {\bibfield  {journal} {\bibinfo  {journal} {Journal of
  Chemical Physics}\ }\textbf {\bibinfo {volume} {131}},\ \bibinfo {pages}
  {054106} (\bibinfo {year} {2009})}\BibitemShut {NoStop}%
\bibitem [{\citenamefont {Booth}\ \emph {et~al.}(2013)\citenamefont {Booth},
  \citenamefont {Gr{\"u}neis}, \citenamefont {Kresse},\ and\ \citenamefont
  {Alavi}}]{booth2013towards}%
  \BibitemOpen
  \bibfield  {author} {\bibinfo {author} {\bibfnamefont {G.~H.}\ \bibnamefont
  {Booth}}, \bibinfo {author} {\bibfnamefont {A.}~\bibnamefont {Gr{\"u}neis}},
  \bibinfo {author} {\bibfnamefont {G.}~\bibnamefont {Kresse}}, \ and\ \bibinfo
  {author} {\bibfnamefont {A.}~\bibnamefont {Alavi}},\ }\href@noop {}
  {\bibfield  {journal} {\bibinfo  {journal} {Nature}\ }\textbf {\bibinfo
  {volume} {493}},\ \bibinfo {pages} {365} (\bibinfo {year}
  {2013})}\BibitemShut {NoStop}%
\bibitem [{\citenamefont {Hall}(2002{\natexlab{a}})}]{hall2002adaptive}%
  \BibitemOpen
  \bibfield  {author} {\bibinfo {author} {\bibfnamefont {R.~W.}\ \bibnamefont
  {Hall}},\ }\href@noop {} {\bibfield  {journal} {\bibinfo  {journal} {Journal
  of Chemical Physics}\ }\textbf {\bibinfo {volume} {116}},\ \bibinfo {pages}
  {1} (\bibinfo {year} {2002}{\natexlab{a}})}\BibitemShut {NoStop}%
\bibitem [{\citenamefont {Hall}(2002{\natexlab{b}})}]{hall2002kink}%
  \BibitemOpen
  \bibfield  {author} {\bibinfo {author} {\bibfnamefont {R.~W.}\ \bibnamefont
  {Hall}},\ }\href@noop {} {\bibfield  {journal} {\bibinfo  {journal} {Chemical
  Physics Letters}\ }\textbf {\bibinfo {volume} {362}},\ \bibinfo {pages} {549}
  (\bibinfo {year} {2002}{\natexlab{b}})}\BibitemShut {NoStop}%
\bibitem [{\citenamefont {Hall}(2005)}]{hall2005simulation}%
  \BibitemOpen
  \bibfield  {author} {\bibinfo {author} {\bibfnamefont {R.~W.}\ \bibnamefont
  {Hall}},\ }\href@noop {} {\bibfield  {journal} {\bibinfo  {journal} {Journal
  of Chemical Physics}\ }\textbf {\bibinfo {volume} {122}},\ \bibinfo {pages}
  {164112} (\bibinfo {year} {2005})}\BibitemShut {NoStop}%
\bibitem [{\citenamefont {Anderson}, \citenamefont {Yuval},\ and\ \citenamefont
  {Hamann}(1970)}]{anderson1970exact}%
  \BibitemOpen
  \bibfield  {author} {\bibinfo {author} {\bibfnamefont {P.~W.}\ \bibnamefont
  {Anderson}}, \bibinfo {author} {\bibfnamefont {G.}~\bibnamefont {Yuval}}, \
  and\ \bibinfo {author} {\bibfnamefont {D.}~\bibnamefont {Hamann}},\
  }\href@noop {} {\bibfield  {journal} {\bibinfo  {journal} {Physical Review
  B}\ }\textbf {\bibinfo {volume} {1}},\ \bibinfo {pages} {4464} (\bibinfo
  {year} {1970})}\BibitemShut {NoStop}%
\bibitem [{\citenamefont {Chiles}\ \emph {et~al.}(1984)\citenamefont {Chiles},
  \citenamefont {Jongeward}, \citenamefont {Bolton},\ and\ \citenamefont
  {Wolynes}}]{chiles1984monte}%
  \BibitemOpen
  \bibfield  {author} {\bibinfo {author} {\bibfnamefont {R.}~\bibnamefont
  {Chiles}}, \bibinfo {author} {\bibfnamefont {G.}~\bibnamefont {Jongeward}},
  \bibinfo {author} {\bibfnamefont {M.}~\bibnamefont {Bolton}}, \ and\ \bibinfo
  {author} {\bibfnamefont {P.}~\bibnamefont {Wolynes}},\ }\href@noop {}
  {\bibfield  {journal} {\bibinfo  {journal} {Journal of Chemical Physics}\
  }\textbf {\bibinfo {volume} {81}},\ \bibinfo {pages} {2039} (\bibinfo {year}
  {1984})}\BibitemShut {NoStop}%
\bibitem [{\citenamefont {Feller}(1996)}]{feller1996role}%
  \BibitemOpen
  \bibfield  {author} {\bibinfo {author} {\bibfnamefont {D.}~\bibnamefont
  {Feller}},\ }\href@noop {} {\bibfield  {journal} {\bibinfo  {journal}
  {Journal of Computational Chemistry}\ }\textbf {\bibinfo {volume} {17}},\
  \bibinfo {pages} {1571} (\bibinfo {year} {1996})}\BibitemShut {NoStop}%
\bibitem [{\citenamefont {Schuchardt}\ \emph {et~al.}(2007)\citenamefont
  {Schuchardt}, \citenamefont {Didier}, \citenamefont {Elsethagen},
  \citenamefont {Sun}, \citenamefont {Gurumoorthi}, \citenamefont {Chase},
  \citenamefont {Li},\ and\ \citenamefont {Windus}}]{schuchardt2007basis}%
  \BibitemOpen
  \bibfield  {author} {\bibinfo {author} {\bibfnamefont {K.~L.}\ \bibnamefont
  {Schuchardt}}, \bibinfo {author} {\bibfnamefont {B.~T.}\ \bibnamefont
  {Didier}}, \bibinfo {author} {\bibfnamefont {T.}~\bibnamefont {Elsethagen}},
  \bibinfo {author} {\bibfnamefont {L.}~\bibnamefont {Sun}}, \bibinfo {author}
  {\bibfnamefont {V.}~\bibnamefont {Gurumoorthi}}, \bibinfo {author}
  {\bibfnamefont {J.}~\bibnamefont {Chase}}, \bibinfo {author} {\bibfnamefont
  {J.}~\bibnamefont {Li}}, \ and\ \bibinfo {author} {\bibfnamefont {T.~L.}\
  \bibnamefont {Windus}},\ }\href@noop {} {\bibfield  {journal} {\bibinfo
  {journal} {Journal of Chemical Information and Modeling}\ }\textbf {\bibinfo
  {volume} {47}},\ \bibinfo {pages} {1045} (\bibinfo {year}
  {2007})}\BibitemShut {NoStop}%
\bibitem [{\citenamefont {Dunning~Jr}(1970)}]{dunning1970gaussian}%
  \BibitemOpen
  \bibfield  {author} {\bibinfo {author} {\bibfnamefont {T.~H.}\ \bibnamefont
  {Dunning~Jr}},\ }\href@noop {} {\bibfield  {journal} {\bibinfo  {journal}
  {Journal of Chemical Physics}\ }\textbf {\bibinfo {volume} {53}},\ \bibinfo
  {pages} {2823} (\bibinfo {year} {1970})}\BibitemShut {NoStop}%
\bibitem [{\citenamefont {Dunning}, \citenamefont {Hay},\ and\ \citenamefont
  {Schaefer}(1977)}]{dunning1977methods}%
  \BibitemOpen
  \bibfield  {author} {\bibinfo {author} {\bibfnamefont {T.~H.}\ \bibnamefont
  {Dunning}}, \bibinfo {author} {\bibfnamefont {P.~J.}\ \bibnamefont {Hay}}, \
  and\ \bibinfo {author} {\bibfnamefont {H.}~\bibnamefont {Schaefer}},\ }in\
  \href@noop {} {\emph {\bibinfo {booktitle} {Modern Theoretical Chemistry}}},\
  Vol.~\bibinfo {volume} {3}\ (\bibinfo  {publisher} {Plenum Press New York},\
  \bibinfo {year} {1977})\ p.~\bibinfo {pages} {1}\BibitemShut {NoStop}%
\bibitem [{\citenamefont {Valiev}\ \emph {et~al.}(2010)\citenamefont {Valiev},
  \citenamefont {Bylaska}, \citenamefont {Govind}, \citenamefont {Kowalski},
  \citenamefont {Straatsma}, \citenamefont {Van~Dam}, \citenamefont {Wang},
  \citenamefont {Nieplocha}, \citenamefont {Apra}, \citenamefont {Windus} \emph
  {et~al.}}]{valiev2010nwchem}%
  \BibitemOpen
  \bibfield  {author} {\bibinfo {author} {\bibfnamefont {M.}~\bibnamefont
  {Valiev}}, \bibinfo {author} {\bibfnamefont {E.~J.}\ \bibnamefont {Bylaska}},
  \bibinfo {author} {\bibfnamefont {N.}~\bibnamefont {Govind}}, \bibinfo
  {author} {\bibfnamefont {K.}~\bibnamefont {Kowalski}}, \bibinfo {author}
  {\bibfnamefont {T.~P.}\ \bibnamefont {Straatsma}}, \bibinfo {author}
  {\bibfnamefont {H.~J.}\ \bibnamefont {Van~Dam}}, \bibinfo {author}
  {\bibfnamefont {D.}~\bibnamefont {Wang}}, \bibinfo {author} {\bibfnamefont
  {J.}~\bibnamefont {Nieplocha}}, \bibinfo {author} {\bibfnamefont
  {E.}~\bibnamefont {Apra}}, \bibinfo {author} {\bibfnamefont {T.~L.}\
  \bibnamefont {Windus}},  \emph {et~al.},\ }\href@noop {} {\bibfield
  {journal} {\bibinfo  {journal} {Computer Physics Communications}\ }\textbf
  {\bibinfo {volume} {181}},\ \bibinfo {pages} {1477} (\bibinfo {year}
  {2010})}\BibitemShut {NoStop}%
\bibitem [{\citenamefont {Werner}\ \emph {et~al.}(2012)\citenamefont {Werner},
  \citenamefont {Knowles}, \citenamefont {Knizia}, \citenamefont {Manby},\ and\
  \citenamefont {Sch{\"u}tz}}]{werner2012molpro}%
  \BibitemOpen
  \bibfield  {author} {\bibinfo {author} {\bibfnamefont {H.-J.}\ \bibnamefont
  {Werner}}, \bibinfo {author} {\bibfnamefont {P.~J.}\ \bibnamefont {Knowles}},
  \bibinfo {author} {\bibfnamefont {G.}~\bibnamefont {Knizia}}, \bibinfo
  {author} {\bibfnamefont {F.~R.}\ \bibnamefont {Manby}}, \ and\ \bibinfo
  {author} {\bibfnamefont {M.}~\bibnamefont {Sch{\"u}tz}},\ }\href@noop {}
  {\bibfield  {journal} {\bibinfo  {journal} {Wiley Interdisciplinary Reviews:
  Computational Molecular Science}\ }\textbf {\bibinfo {volume} {2}},\ \bibinfo
  {pages} {242} (\bibinfo {year} {2012})}\BibitemShut {NoStop}%
\bibitem [{mol()}]{molpro}%
  \BibitemOpen
  \href@noop {} {}\bibinfo {note} {{\sc molpro} is a package of {ab} {initio}
  programs written by H.-J. Werner and P. J. Knowles, with contributions from
  J. Alml{\"{o}}f, R. D. Amos, A. Berning, D. L. Cooper, M. J. O. Deegan, A. J.
  Dobbyn, F. Eckert, S. T. Elbert, C. Hampel, R. Lindh, A. W. Lloyd, W. Meyer,
  A. Nicklass, K. Peterson, R. Pitzer, A. J. Stone, P. R. Taylor, M. E. Mura,
  P. Pulay, M. Sch{\"{u}}tz, H. Stoll, and T. Thorsteinsson.}\BibitemShut
  {Stop}%
\bibitem [{\citenamefont {Saxe}, \citenamefont {Shaefer},\ and\ \citenamefont
  {Handy}(1981)}]{saxe1981exact}%
  \BibitemOpen
  \bibfield  {author} {\bibinfo {author} {\bibfnamefont {P.}~\bibnamefont
  {Saxe}}, \bibinfo {author} {\bibfnamefont {H.~F.}\ \bibnamefont {Shaefer}}, \
  and\ \bibinfo {author} {\bibfnamefont {N.~C.}\ \bibnamefont {Handy}},\
  }\href@noop {} {\bibfield  {journal} {\bibinfo  {journal} {Chemical Physics
  Letters}\ }\textbf {\bibinfo {volume} {79}},\ \bibinfo {pages} {202}
  (\bibinfo {year} {1981})}\BibitemShut {NoStop}%
\bibitem [{\citenamefont {Bash}\ \emph {et~al.}(1996)\citenamefont {Bash},
  \citenamefont {Ho}, \citenamefont {MacKerell}, \citenamefont {Levine},\ and\
  \citenamefont {Hallstrom}}]{bash1996progress}%
  \BibitemOpen
  \bibfield  {author} {\bibinfo {author} {\bibfnamefont {P.~A.}\ \bibnamefont
  {Bash}}, \bibinfo {author} {\bibfnamefont {L.~L.}\ \bibnamefont {Ho}},
  \bibinfo {author} {\bibfnamefont {A.~D.}\ \bibnamefont {MacKerell}}, \bibinfo
  {author} {\bibfnamefont {D.}~\bibnamefont {Levine}}, \ and\ \bibinfo {author}
  {\bibfnamefont {P.}~\bibnamefont {Hallstrom}},\ }\href@noop {} {\bibfield
  {journal} {\bibinfo  {journal} {Proceedings of the National Academy of
  Sciences}\ }\textbf {\bibinfo {volume} {93}},\ \bibinfo {pages} {3698}
  (\bibinfo {year} {1996})}\BibitemShut {NoStop}%
\bibitem [{\citenamefont {Noga}\ and\ \citenamefont
  {Bartlett}(1987)}]{noga1987full}%
  \BibitemOpen
  \bibfield  {author} {\bibinfo {author} {\bibfnamefont {J.}~\bibnamefont
  {Noga}}\ and\ \bibinfo {author} {\bibfnamefont {R.~J.}\ \bibnamefont
  {Bartlett}},\ }\href@noop {} {\bibfield  {journal} {\bibinfo  {journal}
  {Journal of Chemical Physics}\ }\textbf {\bibinfo {volume} {86}},\ \bibinfo
  {pages} {7041} (\bibinfo {year} {1987})}\BibitemShut {NoStop}%
\bibitem [{\citenamefont {Bhaskaran-Nair}\ \emph {et~al.}(2012)\citenamefont
  {Bhaskaran-Nair}, \citenamefont {Brabec}, \citenamefont {Apr{\`a}},
  \citenamefont {van Dam}, \citenamefont {Pittner},\ and\ \citenamefont
  {Kowalski}}]{bhaskaran2012implementation}%
  \BibitemOpen
  \bibfield  {author} {\bibinfo {author} {\bibfnamefont {K.}~\bibnamefont
  {Bhaskaran-Nair}}, \bibinfo {author} {\bibfnamefont {J.}~\bibnamefont
  {Brabec}}, \bibinfo {author} {\bibfnamefont {E.}~\bibnamefont {Apr{\`a}}},
  \bibinfo {author} {\bibfnamefont {H.~J.}\ \bibnamefont {van Dam}}, \bibinfo
  {author} {\bibfnamefont {J.}~\bibnamefont {Pittner}}, \ and\ \bibinfo
  {author} {\bibfnamefont {K.}~\bibnamefont {Kowalski}},\ }\href@noop {}
  {\bibfield  {journal} {\bibinfo  {journal} {Journal of Chemical Physics}\
  }\textbf {\bibinfo {volume} {137}},\ \bibinfo {pages} {094112} (\bibinfo
  {year} {2012})}\BibitemShut {NoStop}%
\bibitem [{\citenamefont {Kowalski}\ and\ \citenamefont
  {Piecuch}(2000)}]{kowalski2000renormalized}%
  \BibitemOpen
  \bibfield  {author} {\bibinfo {author} {\bibfnamefont {K.}~\bibnamefont
  {Kowalski}}\ and\ \bibinfo {author} {\bibfnamefont {P.}~\bibnamefont
  {Piecuch}},\ }\href@noop {} {\bibfield  {journal} {\bibinfo  {journal}
  {Journal of Chemical Physics}\ }\textbf {\bibinfo {volume} {113}},\ \bibinfo
  {pages} {5644} (\bibinfo {year} {2000})}\BibitemShut {NoStop}%
\bibitem [{\citenamefont {Kowalski}\ and\ \citenamefont
  {Piecuch}(2001)}]{kowalski2001comparison}%
  \BibitemOpen
  \bibfield  {author} {\bibinfo {author} {\bibfnamefont {K.}~\bibnamefont
  {Kowalski}}\ and\ \bibinfo {author} {\bibfnamefont {P.}~\bibnamefont
  {Piecuch}},\ }\href@noop {} {\bibfield  {journal} {\bibinfo  {journal}
  {Chemical Physics Letters}\ }\textbf {\bibinfo {volume} {344}},\ \bibinfo
  {pages} {165} (\bibinfo {year} {2001})}\BibitemShut {NoStop}%
\bibitem [{\citenamefont {Mok}, \citenamefont {Neumann},\ and\ \citenamefont
  {Handy}(1996)}]{mok1996dynamical}%
  \BibitemOpen
  \bibfield  {author} {\bibinfo {author} {\bibfnamefont {D.~K.}\ \bibnamefont
  {Mok}}, \bibinfo {author} {\bibfnamefont {R.}~\bibnamefont {Neumann}}, \ and\
  \bibinfo {author} {\bibfnamefont {N.~C.}\ \bibnamefont {Handy}},\ }\href@noop
  {} {\bibfield  {journal} {\bibinfo  {journal} {The Journal of Physical
  Chemistry}\ }\textbf {\bibinfo {volume} {100}},\ \bibinfo {pages} {6225}
  (\bibinfo {year} {1996})}\BibitemShut {NoStop}%
\bibitem [{\citenamefont {Lu}\ \emph {et~al.}(2014)\citenamefont {Lu},
  \citenamefont {H\"oppner}, \citenamefont {Gunnarsson},\ and\ \citenamefont
  {Haverkort}}]{maurits2014}%
  \BibitemOpen
  \bibfield  {author} {\bibinfo {author} {\bibfnamefont {Y.}~\bibnamefont
  {Lu}}, \bibinfo {author} {\bibfnamefont {M.}~\bibnamefont {H\"oppner}},
  \bibinfo {author} {\bibfnamefont {O.}~\bibnamefont {Gunnarsson}}, \ and\
  \bibinfo {author} {\bibfnamefont {M.~W.}\ \bibnamefont {Haverkort}},\ }\href
  {\doibase 10.1103/PhysRevB.90.085102} {\bibfield  {journal} {\bibinfo
  {journal} {Phys. Rev. B}\ }\textbf {\bibinfo {volume} {90}},\ \bibinfo
  {pages} {085102} (\bibinfo {year} {2014})}\BibitemShut {NoStop}%
\end{thebibliography}
\end{document}